\documentclass[structabstract]{aa}  

\usepackage[varg]{txfonts}
\usepackage{graphicx}
\usepackage{rotating}
\usepackage{pdflscape}
\usepackage{lscape}
\usepackage{longtable}
\usepackage{enumerate}
\usepackage[flushleft]{threeparttable}
\usepackage{ulem}

\usepackage{natbib,twoopt}
\bibpunct{(}{)}{;}{a}{}{,}             
\begin{document}

\title{The Gaia-ESO Survey: metallicity of the Chamaeleon~I star forming region\thanks{Based on observations collected at the ESO telescopes under programme 188.B3002, the Gaia-ESO large public spectroscopic survey.}}

\author{L. Spina\inst{1}, S. Randich\inst{1}, F. Palla\inst{1}, K. Biazzo\inst{2}, G.~G. Sacco\inst{1}, E. J. Alfaro\inst{3}, E. Franciosini\inst{1}, L. Magrini\inst{1}, L. Morbidelli\inst{1}, A. Frasca\inst{2}, V. Adibekyan\inst{4}, E. Delgado-Mena\inst{4}, S.~G. Sousa\inst{4},\inst{5}, J.~I. Gonz\'{a}lez Hern\'{a}ndez\inst{6,7}, D. Montes\inst{8}, H. Tabernero\inst{8}, G. Tautvai\v{s}ien\.{e}\inst{9}, R. Bonito\inst{10}, A.~C. Lanzafame\inst{11}, G. Gilmore\inst{12}, R.~D. Jeffries\inst{13}, A. Vallenari\inst{14}, T. Bensby\inst{15}, A. Bragaglia\inst{16}, E. Flaccomio\inst{10}, A.~J. Korn\inst{17}, E. Pancino\inst{16,18}, A. Recio-Blanco\inst{19}, R. Smiljanic\inst{20}, M. Bergemann\inst{12}, M.~T. Costado\inst{3}, F. Damiani\inst{10}, V. Hill\inst{19}, A. Hourihane\inst{12}, P. Jofr\'e\inst{12}, P. de Laverny\inst{19}, C. Lardo\inst{16}, T. Masseron\inst{12}, L. Prisinzano\inst{10}, C.~C. Worley\inst{12}}
\offprints{L. Spina}

\institute{
INAF--Osservatorio Astrofisico di Arcetri, Largo E. Fermi, 5, I-50125 Firenze, Italy \email{lspina@arcetri.astro.it}
\and
INAF--Osservatorio Astrofisico di Catania, via S. Sofia, 78, I-95123 Catania, Italy
\and
Instituto de Astrof\'{i}sica de Andaluc\'{i}a-CSIC, Apdo. 3004, 18080 Granada, Spain
\and
Centro de Astrofisica, Universidade do Porto, Rua das Estrelas, 4150-762, Porto, Portugal
\and
Departamento de F\'isica e Astronomia, Faculdade de Ci\^encias, Universidade do Porto, Rua do Campo Alegre, 4169-007 Porto, Portugal
\and
Instituto de Astrofisica de Canarias (IAC), E-38205 La Laguna, Tenerife, Spain
\and
Depto. Astrofisica, Universidad de La Laguna (ULL), E-38206 La Laguna, Tenerife, Spain
\and
Departamento de Astrofisica, Universidad Complutense de Madrid (UCM), Spain
\and
Institute of Theoretical Physics and Astronomy, Vilnius University, Gostauto 12, Vilnius LT-01108, Lithuania
\and
INAF - Osservatorio Astronomico di Palermo, Piazza del Parlamento 1, 90134, Palermo, Italy
\and
Dipartimento di Fisica e Astronomia, Sezione Astrofisica, Universit\`{a} di Catania, via S. Sofia 78, 95123, Catania, Italy
\and
Institute of Astronomy, University of Cambridge, Madingley Road, Cambridge CB3 0HA, United Kingdom
\and
Astrophysics Group, Research Institute for the Environment, Physical Sciences and Applied Mathematics, Keele University, Keele, Staffordshire ST5 5BG, United Kingdom
\and
INAF - Padova Observatory, Vicolo dell'Osservatorio 5, 35122 Padova, Italy
\and
Lund Observatory, Department of Astronomy and Theoretical Physics, Box 43, SE-221 00 Lund, Sweden
\and
INAF - Osservatorio Astronomico di Bologna, via Ranzani 1, 40127, Bologna, Italy
\and
Department of Physics and Astronomy, Uppsala University, Box 516, SE-75120 Uppsala, Sweden
\and
ASI Science Data Center, Via del Politecnico SNC, 00133 Roma, Italy
\and
Laboratoire Lagrange (UMR7293), Universit\'e de Nice Sophia Antipolis, CNRS,Observatoire de la C\^ote d'Azur, CS 34229,F-06304 Nice cedex 4, France
\and
Department for Astrophysics, Nicolaus Copernicus Astronomical Center, ul. Rabia\'{n}ska 8, 87-100 Toru\'{n}, Poland
}
 \date{Received 5 May 2014; Accepted 10 June 2014}
 \abstract
{Recent metallicity determinations in young open clusters and star-forming regions suggest that the latter may be characterized by a slightly lower metallicity than the Sun and 
older clusters in the solar vicinity. However, these results are based on small statistics and inhomogeneous analyses. The Gaia-ESO Survey is observing and homogeneously analyzing large samples of stars in several young clusters and star-forming regions, hence allowing us to further investigate this issue.} 
{We present a new metallicity  determination
of the Chamaeleon~I star-forming region, based on the products distributed
in the first internal release of the Gaia-ESO Survey.
}
{48 candidate members of Chamaeleon~I have been observed with the high-resolution spectrograph
UVES. We use the surface gravity, lithium line equivalent width and position in the Hertzsprung-Russell diagram to confirm the cluster members and we use the iron abundance to derive the mean metallicity of the region.}
{
Out of the 48 targets, we confirm 15 high probability 
members. Considering the metallicity measurements for 9 of them, we 
find that the iron abundance of Chamaeleon~I is slightly subsolar with a mean value [Fe/H]$=-0.08\pm0.04$~dex. This result is in agreement with the metallicity determination of other nearby star-forming regions and suggests that the chemical pattern of the youngest stars in the solar neighborhood is indeed more metal-poor than the Sun. We argue that this evidence may be related to the chemical distribution of the Gould Belt that contains most of the nearby star-forming regions and young clusters.
}
{}
\authorrunning{L. Spina et al.}
\titlerunning{The Gaia-ESO Survey: metallicity of the Chamaeleon I star-forming region}
\maketitle
\section{Introduction}
The metallicity determination in young open clusters (YOCs) and star-forming regions (SFRs) has implications for fundamental topics, such as the origin and early evolution of these environments, the evolution of circumstellar disks, and the ability to form planets (see \citealt[and references therein]{Gilli06,Neves09,Ercolano10,Yasui10,Kang11,Spezzi12,Adibekyan12a,Adibekyan12b,Spina14}). Furthermore, these young regions are of particular interest since they are still close to their birthplace and contain a homogeneous stellar population that had no time to disperse through the Galactic disk. Thus, YOCs and SFRs are key objects to trace the present chemical pattern of the Galactic thin disk. Indeed, in the last few years an increasing number of studies has focused on the metallicity of YOCs and SFRs (e.g., \citealt{James06,GonzalezHernandez08,Santos08,DOrazi09a,DOrazi09b,DOrazi11,VianaAlmeida09,Biazzo11a,Biazzo11b,Biazzo12a,Biazzo12b,Spina14}). 
These studies suggest that YOCs, where star formation has ceased,
generally share
a metallicity close to the solar value; on the other hand, SFRs, 
in which the molecular gas is still present and the star formation process is still ongoing, seem -surprisingly- 
to be characterized by a somewhat lower iron content
.  Thequestion arises whether this result is
due to low-number statistics and/or to inhomogeneous methods to derive 
the metallicity or to the fact that the metallicity determination in very young stars
in SFRs is more uncertain and generally based on rather cool stars. 
On the other hand, if confirmed, this result would have
important implications for our understanding of the origin of YOCs
and SFRs.

The Gaia-ESO Survey \citep{Gilmore12,Randich13} is observing a significant number of young environments. Whilst the main goal of the young cluster observations
is the study of
their kinematics and dynamical evolution through the measurement of accurate
radial velocities (e.g., \citealt{Jeffries14}), this large amount of data can also be used to perform an homogeneous study of the elemental abundances of YOCs and SFRs. In this framework, in a recent study we have determined the metal content of Gamma Velorum, the first YOC observed by the Gaia-ESO Survey \citep{Spina14}. The present paper is devoted to the analysis of the metallicity of the first SFR targeted by the Gaia-ESO Survey: Chamaeleon~I (hereafter, Cha~I). A detailed analysis of the membership and other properties of the cluster, mostly based on the GIRAFFE data, will be reported in \citet{Sacco14b}.

With a mean age of $\sim$2~Myr \citep{Luhman07} and its proximity to the Sun (d=160-165 pc; \citealt{Whittet97}), Cha~I is one of the best studied SFRs. It is part of a wider star-forming complex, distributed over a region of a few square degrees, that contains also two smaller molecular clouds, Cha~II and Cha~III \citep{schwartz77}. Cha~I has been the target of many spectroscopic and photometric surveys that have uncovered a large population of embedded and optically visible sources (see the review by \citealt{Luhman08}; hereafter, L08). The current sample of known members comprises 237 sources (hereafter, ``L08-mem''), extending down to substellar objects. The census is nearly complete in the central regions of Cha~I (1$1^{\circ}$ 05' $\le$ RA $\le $1$1^{\circ}$ 11'; $-$7$7^{\circ}$ 48' $\le$ DEC $\le$ $-$7$6^{\circ}$ 18') for M$\ge$0.03$M_{\sun}$ and $A_{J}$$\le $1.2, but outside this area the stellar population is still not completely identified \citep{Luhman07}. Recently, \citet{Lopezmarti13} have identified 51 new kinematical candidate members that await confirmation through accurate spectroscopic data. The Initial Mass Function (IMF) of Cha~I has been explored down to substellar masses by \citet{Luhman07} and, as with other SFRs, it reaches a maximum between 0.1 and 0.2~$M_{\sun}$. L08, using {\it Spitzer} colors to study the disk population, argued that the lifetimes of disks around solar-mass stars are longer in Cha~I than in other young clusters, probably because of the lower stellar density and resulting reduction in dynamical interactions. On the other hand, Cha~I is also characterized by a number of subsolar-mass stars with unusually short disk lifetimes \citep{Luhman07,Robberto12}. The Cha~I association can be distinguished in two sub-clusters, Cha~I~North (DEC$>$$-$7$7^{\circ}$) and Cha~I~South (DEC$<$$-$7$7^{\circ}$), with different star formation histories. The distribution of isochronal ages suggests that star formation began $\sim$5-6 Myr ago in the northern portion and developed later in the southern extension \citep{Luhman07}. 

Cha~I is relatively isolated from other SFRs and does not contain numerous massive stars. Two  metallicity determinations have been already published prior to our study: in the
first one \citet{Padgett96} derived an average value of
$<$[Fe/H]$>$~$=$~$-$0.07$\pm$0.06~dex; a subsequent study by
\citet{Santos08} reports $<$[Fe/H]$>$~$=$~$-$0.11$\pm$0.14, but 
this estimate is based on the analysis of four stars that are
located in a wide area of the Chamaeleon complex, away from the main SFRs. Furthermore, in a recent work on proper motions \citet{Lopezmarti13} have shown that two of these four stars, namely RX~J1158.5$-$7754a and RX~J1159.7$-$7601, seem to be kinematical members of the $\epsilon$~Cha association, another one (RX~J1140.3$-$8321) of the $\eta$~Cha association, while RX~J1233.5$-$7523 is a field star. Thus, the metallicity of \citet{Santos08} is not representative of the star-forming region and a new dedicated study is both necessary and timely.


The paper is organized as follows: in Section~$\ref{pipeline}$ we describe the target selection and spectral analysis. The identification of the cluster members on the basis of the surface gravity, the detection of lithium in the stellar atmospheres and their position in the Hertzsprung-Russell diagram is presented in Section~$\ref{mem-analysis}$. The results of the elemental abundance determination are discussed in Section~$\ref{metalcontent}$. In Section \ref{Discussion} we overview and discuss in a broader context the metal content of SFRs. 
Section~$\ref{Conclusions}$ summarizes our findings.

\section{Gaia-ESO data\label{pipeline}}
The analysis presented in this paper is based on the spectroscopic data obtained by the Gaia-ESO Survey during the first six months of observations (January-June 2012) and following the analysis released internally to the survey consortium in the iDR1 catalog at the Wide Field Astronomy Unit at Edinburgh University$\footnote{The GESviDR1Final catalog at http://ges.roe.ac.uk/}$. In this section we describe the target selection, the observations and the available data products of the Gaia-ESO Survey analysis.

\subsection{Target selection and observations} 
The Gaia-ESO Survey observations are performed with the multi-object optical spectrograph FLAMES at the VLT \citep{Pasquini02}, using both GIRAFFE and UVES. 
In this paper we will focus on the latter, while GIRAFFE targets and their
properties will be discussed in a forthcoming paper by \citet{Sacco14b}.

\begin{figure}
\centering
\includegraphics[width=0.5\textwidth]{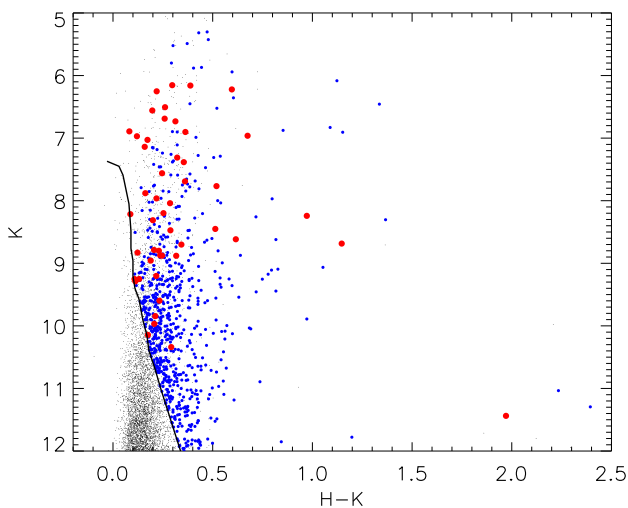}
\caption{Infrared color-magnitude diagram for the stars lying in the cluster field and having $R_{USNO}$$\le$17.0. The black line represents the 10~Myr isochrone using \citet{Siess00} models. Blue (small) and red (large) dots are the stars targeted with GIRAFFE  and UVES, respectively.}
\label{HKcmd_ISO1}
\end{figure}

The selection criteria are based on homogeneous photometric data, covering a large area of the cluster field, following the Gaia-ESO Survey guidelines for cluster observations (see \citealt{Bragaglia14}). UVES targets have been selected
including only those sources: 
i) located in the central and most populated area of the cluster, but wide enough to extend into the cluster boundaries (i.e.,1$0^{\circ}$ 45' $\le$ RA $\le$ 1$1^{\circ}$ 30'; $-$7$9^{\circ}$ 00' $\le$ DEC $\le$ $-$7$5^{\circ}$ 00'); ii) with $R_{USNO}$$\footnote{USNO-A2.0 \citep{Monet08} is a catalog of 526,280,881 stars, that lists right ascension and declination (J2000) and the standard Landolt B and R magnitude for each star.}$$\le$17.0; iii) 
 with avaliable 2MASS photometry \citep{Cutri03}; iv) that fall above the 10~Myr isochrone in the K vs. H-K diagram shown in Fig.~\ref{HKcmd_ISO1} using the \citet{Siess00} models. For the UVES targets high priority was given to stars already identified
as members by L08 with spectral type earlier than $\sim$~M0. However,
several other sources were actually observed for a best exploitation of the
available fibers.

\begin{figure}
\centering
\includegraphics[width=0.5\textwidth]{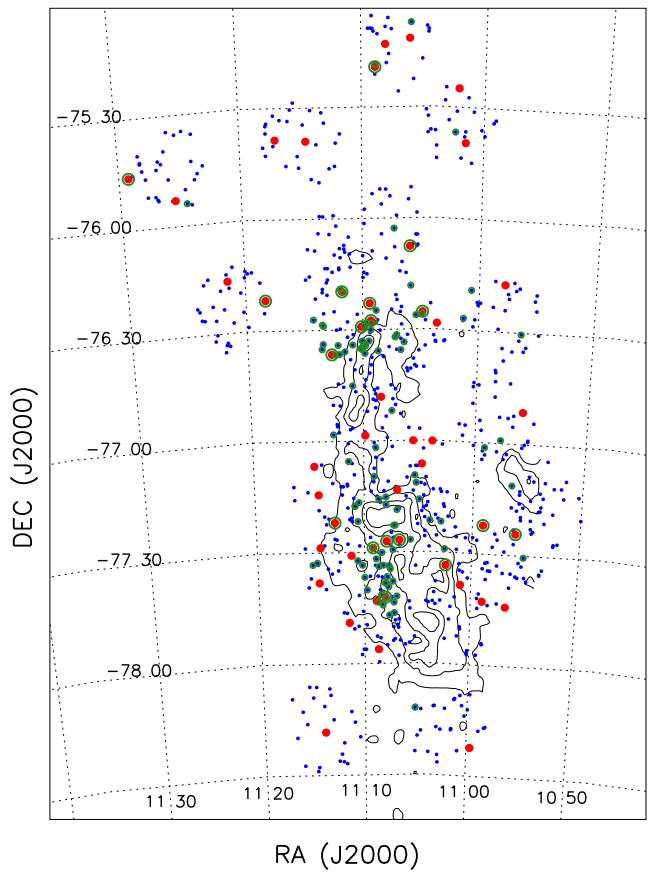}
\caption{Map of the observed sources in the Cha~I fields. Blue (small) dots are the GIRAFFE targets and red (large) dots are the UVES ones. Green circles mark the  ``L08-mem''. The contours correspond to the extinction map of \citet{Cambresy98} at inside-out intervals of $A_{V}$$=$8, 6, 4 and 2.}
\label{MAP_CHA_tot}
\end{figure}

A total of 25 fields covering the regions of Cha~I, as shown in Fig.~\ref{MAP_CHA_tot}, were observed in runs C, D and E (March - May, 2012), using the UVES/CD$\#$3 cross disperser ($\lambda=$4770-6820 $\AA$; R$=$47000). 
Seventeen fields have been chosen in order to cover the central region of the cluster (hereafter, on-fields), characterized by a higher extinction and rich in confirmed members. In order to obtain a complete sampling of the members and possibly discover other candidates missed in previous studies, eight additional fields (off-fields) have been placed in the northern and southern periphery of the association in order to observe its sparse population. Sixteen and nine OBs were observed for 20 ($R_{USNO}$ between 12 and 14 mag)
and 50~min ($R_{USNO}$ between 14 and 17 mag), respectively. 
Eleven of the stars have a longer exposure time because of the partial overlap of the fields. 

A total of 48 UVES spectra were acquired. The sample includes only 18 L08
members as most of the 237 members are M-type stars (and were thus observed with GIRAFFE) or brown dwarfs fainter than the survey limit. One of the kinematic candidate members identified by \citet{Lopezmarti13}, 2MASS~J10593816$-$7822421, 
has also been observed. 
The Signal-to-Noise Ratio (SNR) of the spectra is in the range 5-300, with a median value $\sim$60. The 48 targets are listed in Table~\ref{members} where we include
a running ID, the CNAME, coordinates (J2000), SNR, J magnitude and the membership flag from L08 (``Y'' for members and ``---'' for the stellar object with an unassessed membership from Luhman's papers) and a multiplicity
flag.

\subsection{Available data from the Gaia-ESO Survey\label{Availabledata}}
As mentioned above, we use the products released in the iDR1 catalog
that for the Cha~I region consist of radial velocities, projected rotational 
velocities, spectroscopic cross-correlation functions (CCFs), fundamental stellar
parameters ($\rm T_{\rm eff}$, $\log$~g, [Fe/H]), 
equivalent width of the Li line at 6708~$\AA$ and H$\alpha$, each quantity with its uncertainty. 
The UVES data are reduced using the FLAMES-UVES ESO public pipeline.
The determination of radial and rotational velocities is described
in detail in \citet{Sacco14a}. A specific working group of the Gaia-ESO consortium 
is dedicated to the analysis of cool young stars.For UVES spectra this working group benefit of the contribution of four nodes that use different 
methods of analysis, that can be summarized as follows: i) the equivalent width (EW) analysis: the atmospheric parameter determination is based on the excitation and ionization balance of the iron lines; ii) spectral classification and estimated atmospheric parameters from a $\chi^{2}$ fit of the observed spectra with a grid of templates composed by observed spectra of slow-rotating, low-activity stars. The parameters released in iDR1 catalog are obtained by computing the median value of the results provided by the nodes, after the outliers have been discarded. 
Uncertainties are the node-to-node dispersions. 
We mention that the all the working groups of the consortium devoted to the spectroscopic analysis of F-,G-,K-,M-type stars uniformly makes use of MARCS models of stellar atmospheres \citep{Gustafsson08}, that assume the solar abundances from \citet{Grevesse07}. Also, common atomic data have been used for the analysis of all the spectra of the Gaia-ESO Survey \citep{Heiter14}. 
Similarly, more than one node measure the
strength of the Li~{\sc i} line at 6707.8~$\AA$ in both Giraffe 
and UVES spectra. 
They use independent methods to derive the EW of this features: 
specifically, some of them apply a Gaussian fitting to the line, while others
are based on the direct profile integration of the line. 
The median value of the EW (or the average, when only two nodes provided 
the measurement) corrected for the spectral veiling are then adoped. 
All these procedures are detailed in Lanzafame et al. (in prep).

The available products for the 48 targets are listed in Table~\ref{67parameters} where we report the following quantities: running number, radial velocity, rotational velocities,
fundamental parameters, equivalent width of the Li line, along with the estimate of the bolometric luminosity ($L_{bol}$) and the information on binarity and membership resulting from our analysis
(see Section \ref{mem-analysis}). 
The ${\rm L_{\rm bol}}$ values have been derived from the $J_{\rm 2MASS}$ magnitudes corrected for the extinction and assuming the distance of the cluster being 160~pc as previously determined by \citet{Whittet97}. Namely, the extinction has been estimated from the difference between the photometric and spectroscopic temperatures, while photometric temperatures and bolometric corrections have been derived adopting the calibrations of \citet{Pecaut13}. The ${\rm L_{\rm bol}}$ errors take into account the uncertainties on the magnitudes, spectroscopic temperatures and cluster distance. 
The mean uncertainties on the stellar parameters are: $<$$\sigma_{\rm Teff}$$>$$=$126~K, $<$$\sigma_{\rm log~g}$$>$$=$0.25~dex, $<$$\sigma_{\rm [Fe/H]}$$>$$=$0.13~dex.

As indicated in the table, the values of the main parameters have been 
derived for 42 of the initial 48 UVES targets.
The remaining stars could not be analyzed due to the poor SNR or because of the presence of strong spectral veiling. Radial velocities are available
for 42 targets. Sixteen out of 42 are
L08 members.
Since the main aim of this paper is to determine the metal content of Cha~I, in the following we consider only those stars with the main parameters available.

\section{Membership analysis of UVES targets\label{mem-analysis}}
We have identified two double-lined binaries (SB2) through the spectral CCFs: $\#$22 and $\#$27 in Table~\ref{67parameters}. Their binarity was known also from previous studies (i.e., \citealt{Covino97,Lafreniere08}). We will not consider these systems for membership analysis, even though the Gaia-ESO Survey provides the stellar parameters for one of them, because the determination is likely to be unreliable.

Following the same procedure adopted in \citet{Spina14} for the Gamma Velorum cluster, in this Section we use the spectroscopic information, along with the position of the stars in the Hertzsprung-Russell diagram (HRD), to carry out the membership analysis. This is performed on the 41 UVES targets whose main parameters have been determined by the Gaia-ESO consortium and that have not been flagged as SB2. This sample contains 15 ``L08 mem''.

\subsection{Identification of the giant contaminants \label{logg}}
The sequence of Cha~I members is not clearly identifiable in the CMD of Fig.~\ref{HKcmd_ISO1} since the sample of targets is contaminated by field stars. In order to discard the population of evolved contaminants, we will consider for further analysis only those stars with a spectroscopic $\log~g$$\gtrsim$3.5~dex since stars with lower values of surface gravities are likely background giants. Among the 41 stars with $\log~g$ determinations, 20 have been rejected as evolved contaminants and, as expected, none of them is a ``L08-mem''. On the other hand, all the remaining 21 targets are flagged as candidate 
members and their membership has been assessed using the lithium equivalent width, as shown below. 

\subsection{Lithium members \label{Li-mem}}
\begin{figure}
\centering
\includegraphics[width=0.5\textwidth]{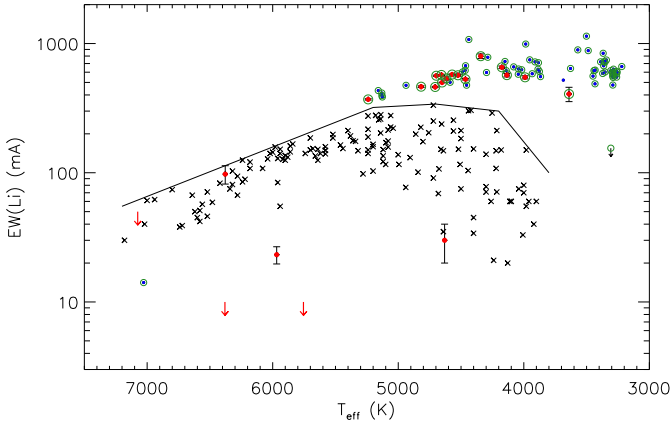}
\caption{Lithium EW as a function of the $\rm T_{\rm eff}$ for the candidate members of Cha~I. The red (large) dots identify the UVES stars. Most of the Li detections in UVES spectra have uncertainties associated to their EWs smaller than the the data points. The GIRAFFE members identified by \citet{Sacco14b} with a $\rm T_{\rm eff}$ determination are shown by the smaller blue dots. All the ``L08-mem'' are marked with a green circle. The solid line denotes the upper-envelope of the Pleiades distribution (crosses; \citealt{Soderblom93,Jones96b}).}
\label{Teff_Li}
\end{figure}

In Fig.~$\ref{Teff_Li}$ we show the EWs as a function of $\rm T_{\rm eff}$ for the 21 UVES sources that have not been rejected as SB2 or giant contaminants. 
In order to identify the sequence of cluster members, we also plot the GIRAFFE
targets confirmed as cluster members by \citet{Sacco14b} on the basis of the equivalent with of the Li line
at 6078 \AA~ and the H$\alpha$ width at the 10\% of the peak. For these stars they list the EWs(Li) and main parameters recommended by the Gaia-ESO Survey, that are the values plotted in Fig.~$\ref{Teff_Li}$.

The distribution of ``L08-mem'', most of which are cooler than  $\sim5300$~K and have a EW(Li)$>$300~m$\AA$, clearly defines the sequence of Li undepleted members. In order to assess the membership of the UVES sources 
on the basis of the lithium content, we also use the available information 
for the members of the Pleiades cluster ($\sim$125-130~Myr; \citealt{Stauffer98}), similarly to
the approach of \citet{Spina14}. The comparison of the EWs(Li) of our stars with those of Pleiades members with
similar $\rm T_{\rm eff}$ will allow us to identify the youngest targets, 
therefore the likely members of Cha~I.
Among the UVES targets
15 stars have EW(Li) higher than their Pleiades counterparts, 
since they lie above the upper-envelope of the Pleiades Li-temperature distribution. Not surprisingly, all these stars, hereafter flagged as Li-members, have EW(Li)$>$300~m$\AA$ and are ``L08-mem''.
{\it Viceversa}, four UVES targets have EW(Li)$\leq$30~m$\AA$ and are located significantly below the Pleiades distribution. These latter are likely field contaminants. Two additional objects warmer than 6000~K, at $\rm T_{\rm eff} =$ 6378~K and 7075~K, lie slightly below 
the upper-envelope of the Pleiades and have lithium EWs which are compatible with a pre-main-sequence cluster. However, since Li is no longer a good
age tracer for such stars, we will consider them as ``hot-candidate-members'' (HCMs) and will try to derive their association with Cha~I from the position in the HRD.

\subsection{Hertzsprung-Russell diagram\label{HRD-mem}}

\begin{figure}
\centering
\includegraphics[width=0.5\textwidth]{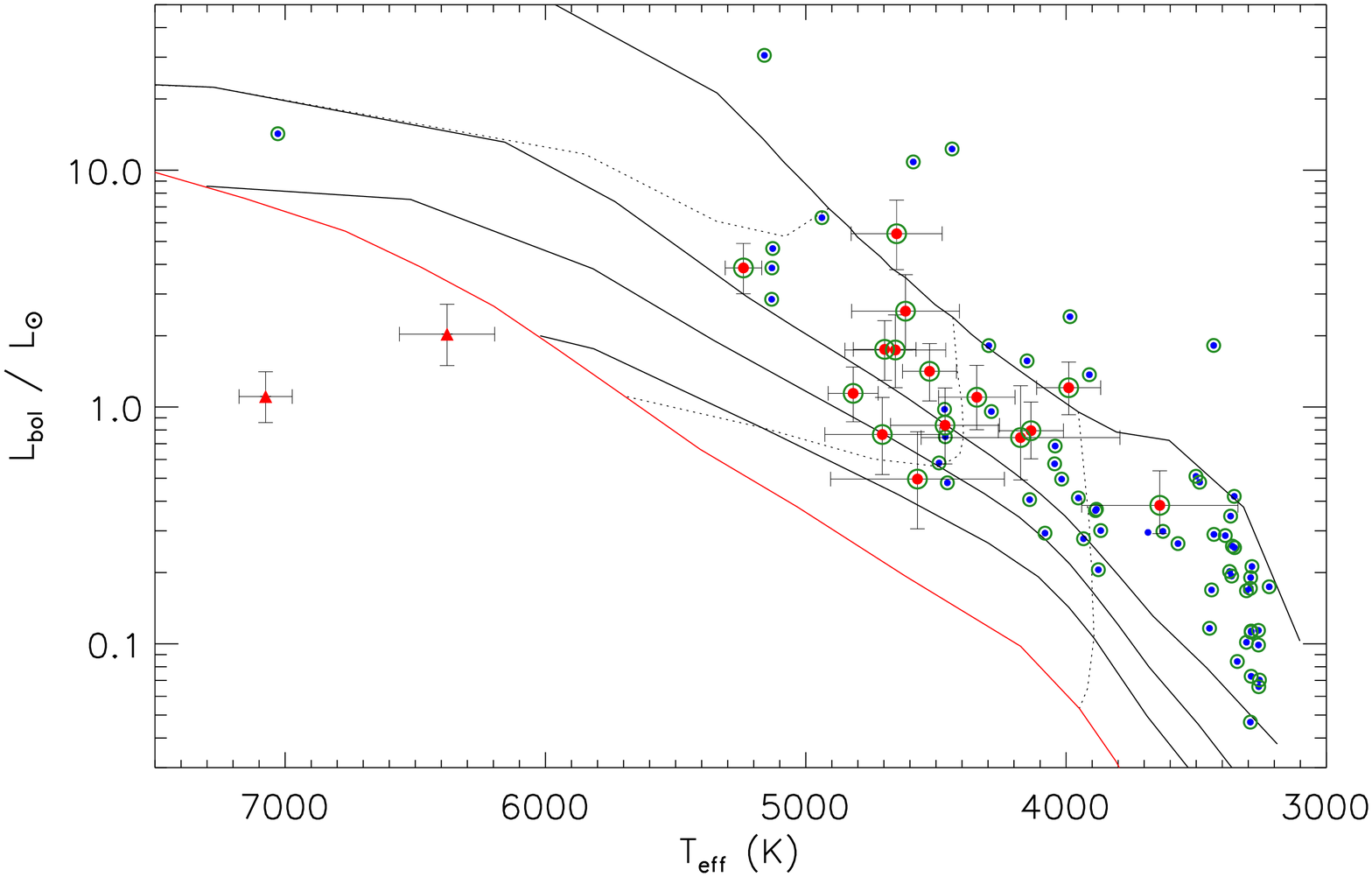}
\caption{HR diagram of the UVES Li-members (red large dots) and HCMs (red triangles). All the ``L08 mem'' are marked with a green circle. The members targeted by GIRAFFE are shown as small blue dots. The dotted and solid black lines are the evolutionary tracks for 0.5, 1 and 2 $M_{\sun}$ and isochrones for 1, 5, 10, 20~Myr, respectively. The ZAMS is marked with a solid red line. The evolutionary tracks, the isochrones and the ZAMS are from the stellar models of \citet{Siess00} for a chemical composition with $Z=$0.01.}
\label{HRD}
\end{figure}

The HRD can be used to test the reliability of our membership analysis and to provide some additional information about the HCMs for which we were not able to establish a secure membership based on lithium. 
Using the bolometric luminosity and the effective temperature from the Gaia-ESO Survey
we plot the UVES Li-members and the HCMs in the HRD of Fig. \ref{HRD}, together with the GIRAFFE members identified by \citet{Sacco14b}. As for the $\rm T_{\rm eff}$, the $\rm L_{bol}$ values adopted for the GIRAFFE targets are those listed by \citet{Sacco14b} and derived with the same procedure used in the present paper for the UVES targets.  Overlaid on the data are the 1, 5, 10 and 20 Myr isochrones, the zero-age-main-sequence (ZAMS) and the evolutionary tracks for stars with 0.5, 1 and 2 $M_{\sun}$ from \citet{Siess00} models for a stellar metallicity of $Z$=0.01. The great majority of the Li-members ($\rm T_{\rm eff}$$<$5500~K) occupy a region of the diagram between the 1 and 5 Myr isochrones, in agreement with the cluster mean age of $\sim$2~Myr estimated by \citet{Luhman07}. On the other hand, both HCMs are considerably below the ZAMS and one of them has
a radial velocity inconsistent with the mean RV for Cha~I reported by \citet{Sacco14b} ($<$RV$>$$=$14.85$\pm$0.018 and $\sigma_{\rm RV}$$=$1.1$\pm$0.16 km/s), suggesting that they are likely contaminants.
Therefore, we reject them from further analysis.

\subsection{Conclusion on the membership analysis \label{GVmem}}
The membership flags for each of the UVES stars derived from $\log~g$, lithium EWs and HRD are summarized
in Table~\ref{67parameters}.
In total, there are 15 secure members that satisfy our membership criteria.
Since all these stars have been previously defined as Cha~I members by L08, there is no new member among the UVES targets. Conversely, all the L08 with Gaia-ESO parameters are confirmed as members. Finally, we have also found that the kinematic candidate member $\#$7 identified by \citet{Lopezmarti13} have a photospheric lithium and a surface gravity incompatible for a pre-main--sequence star, thus it is a likely contaminant.

\begin{figure}
\centering
\includegraphics[width=0.5\textwidth]{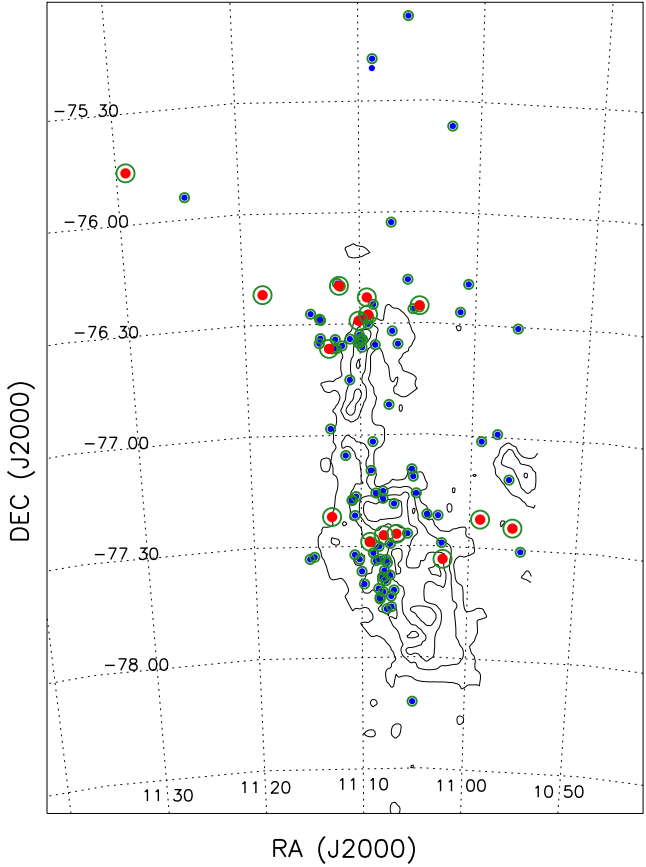}
\caption{Map of the spatial location of the Cha~I members. The symbols are the same as in Fig.~\ref{HRD}. The contours correspond to the extinction map of \citet{Cambresy98} at inside-out intervals of $A_{V}$$=$8, 6, 4 and 2.}
\label{MAP_CHA_mem}
\end{figure}

In Fig.~\ref{MAP_CHA_mem} we show the spatial distribution of the 15 UVES and 103 GIRAFFE members for the Cha~I cloud. We will use this information to check if a difference in metallicity is present between the northern and southern sub-clusters and the distributed population.

\section{The metallicity of Cha~I \label{metalcontent}}

\begin{figure}
\centering
\includegraphics[width=0.5\textwidth]{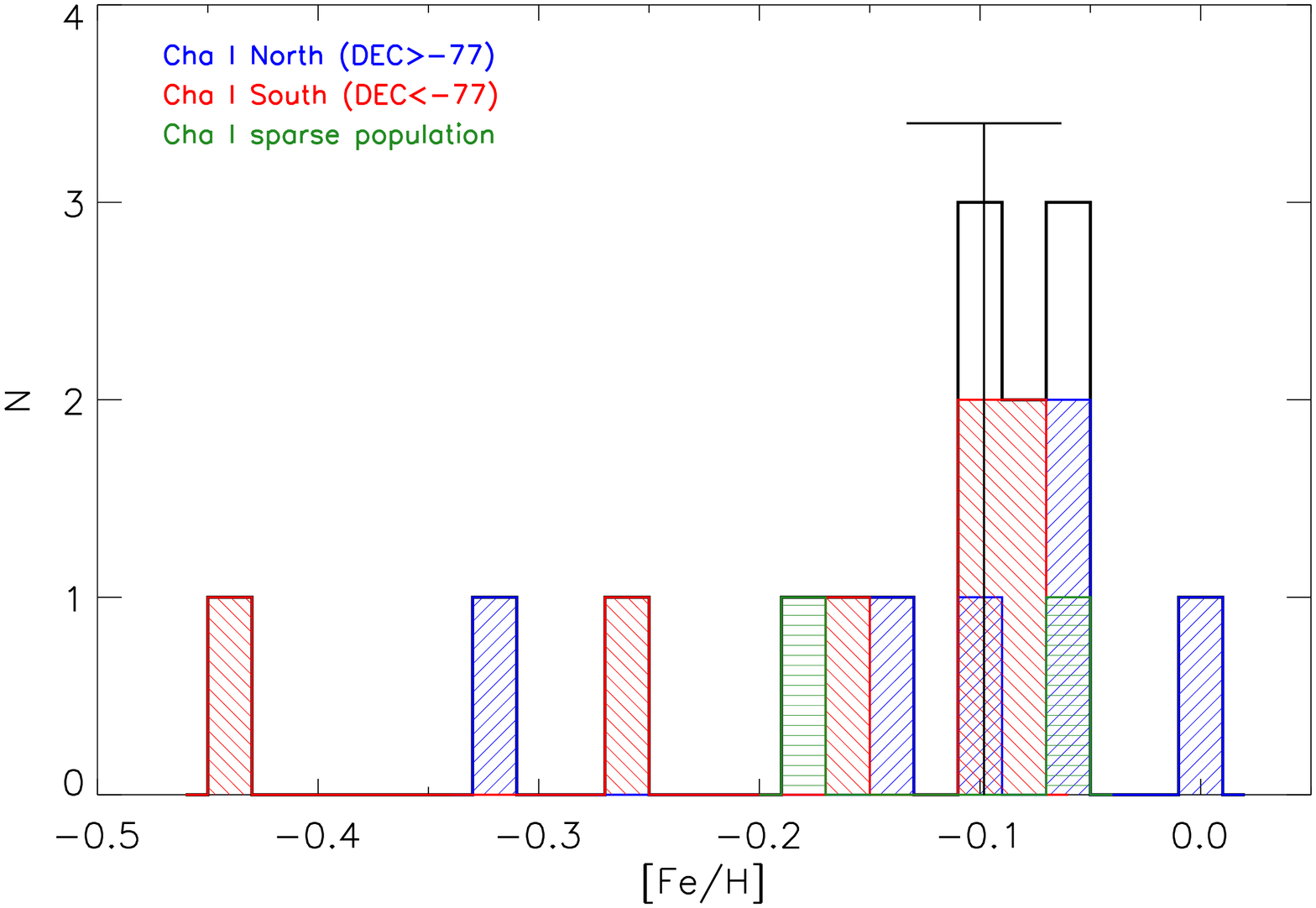}
\caption{Iron abundance distribution of the 15 UVES members. The resulting weighted mean metallicity of Cha~I is $<$[Fe/H]$>$$=$$-$0.10$\pm$0.04 dex. The different colors highlight the contribution of the different populations of the cluster, Cha~I North, South and the sparse population.}
\label{iron_distr}
\end{figure}

Based on the 15 UVES members, we derive the [Fe/H] distribution of Cha~I  shown in Fig.~\ref{iron_distr}. 
The weighted mean of the distribution is $<$[Fe/H]$>$$=$$-$0.10$\pm$0.04 dex. We note that the error of the mean (0.04 dex) is small compared to the width of the distribution that extends from $-$0.45 to 0.00~dex and to the standard deviation $\sigma_{\rm [Fe/H]}$$=$0.13 dex computed without assigning any weight. We believe that the large excursion of [Fe/H] values is due to the lower accuracy of the more discrepant values rather than to a real dispersion. In support of this claim, we observe that the two most-metal poor stars ($\#$28 and 32) are also those with the biggest errors. Furthermore, in Fig.~\ref{iron_distr} we highlight with different colors the results for different regions of the cluster: Cha~I North, Cha~I South and the sparse population. We see that there is no spatial segregation of the iron content in the cluster which is in fact characterized by a rather homogeneous distribution. 

\begin{figure}
\centering
\includegraphics[width=0.5\textwidth]{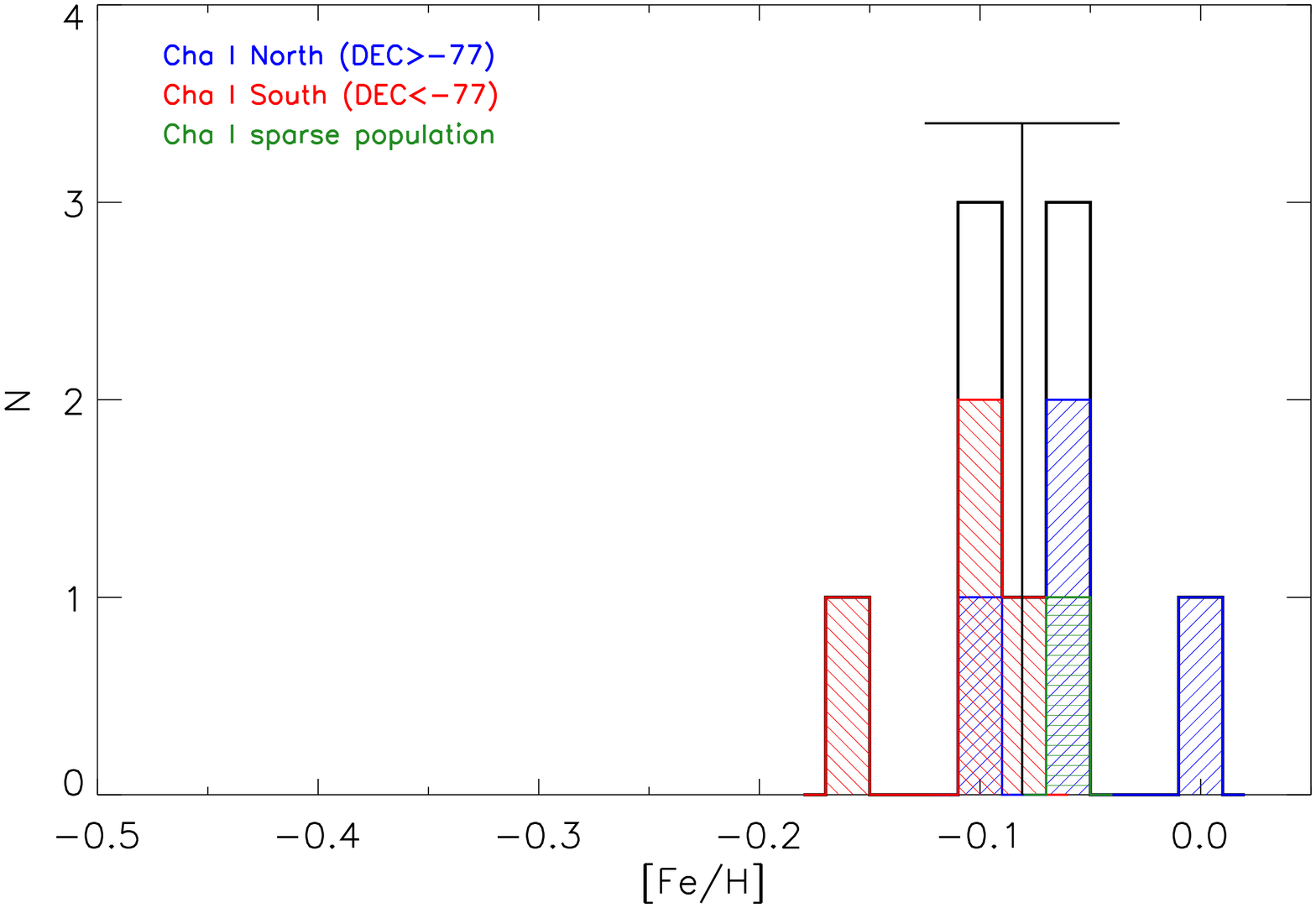}
\caption{Iron abundance of the 9 UVES members that have not been identified as tight binaries by \citet{Lafreniere08}. The resulting weighted mean metallicity of Cha~I is $<$[Fe/H]$>$$=$$-$0.08$\pm$0.04 dex. Colors are the same of Fig. \ref{iron_distr}}
\label{iron_distr_clean}
\end{figure}

Interestingly, six UVES members have been identified by \citet{Lafreniere08} as part of tight$\footnote{Since the UVES fiber has a diameter of 1$\arcsec$, we consider ``tight systems'' only those with a separation $\leq$2$\arcsec$.}$ multiple systems (see column 5 in Table~\ref{members}), namely $\#$21, 28, 29, 32, 36 and 45. Although we do not infer any evidence of binarity from their spectra, if we exclude their [Fe/H] values, we obtain the distribution shown in Fig~\ref{iron_distr_clean}. In this case, the resulting weighted mean is similar to the previous one, $<$[Fe/H]$>$$=$$-$0.08~dex, but with a significantly lower standard deviation $\sigma_{\rm [Fe/H]}$$=$0.04~dex. We will consider this mean as the final metallicity value of Cha~I and take as its error the standard deviation around the mean: [Fe/H]$=$$-$0.08$\pm$0.04~dex.

\begin{figure}
\centering
\includegraphics[width=0.5\textwidth]{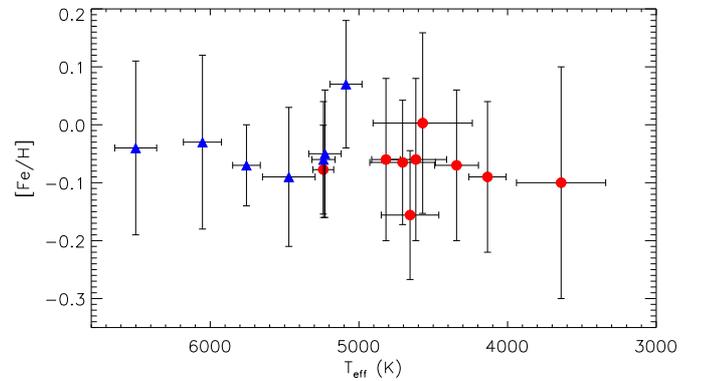}
\caption{Iron abundance as a function of $\rm T_{\rm eff}$ for the nine Cha~I members (red circles) and the seven stars classified as members of Gamma Velorum by \citet{Spina14}}
\label{teff_Fe}
\end{figure}

As a check in order to asses the reliability of the iron abundances produced by the Gaia-ESO Survey, we show in Fig.~\ref{teff_Fe} [Fe/H] as a function of the effective temperature for the nine stars that have been used to derive the mean cluster metallicity. For comparison we also include the results obtained in Gamma Velorum within the Gaia-ESO Survey. Clearly, no trend with temperature is visible across the temperature range $\sim 3500-6500$~K, indicating that no bias is affecting our analysis even at the lowest temperatures.

\section{Discussion \label{Discussion}}

\subsection{Comparison with a previous metallicity determination of CS~Cha}

As described in Section~\ref{metalcontent}, our iron abundance analysis is based on a sample 
of 9 stars observed with UVES. Of these, only one (namely, CS~Cha; \#11 in Table~\ref{members}), 
had been previously observed and analyzed to derive the iron abundance by \cite{Padgett96}. 
The Gaia-ESO value for this star is [Fe/H]=$-0.16\pm0.11$~dex that is very different from that [Fe/H]=$+0.11\pm0.14$~dex quoted by \cite{Padgett96} . 
We believe that the difference comes from the fact that the analysis by Padgett was based on few iron lines (16) compared to the $\sim$100-200 lines generally used in the Gaia-ESO Survey analysis. Moreover, strong lines that are heavily affected by the treatment of damping were not excluded in Padgett's list. In the case of CS~Cha, 5 of the 16 lines used for the iron abundances have EW$>$150~m\AA. As a consequence, the micro-turbulent velocity found by \cite{Padgett96}, $\xi=0.3\pm0.6$ km/s, is
much lower than the mean value of $\xi \sim 1.7$ km/s obtained for the other four stars 
and the similar value found for other young stars by, e.g., \citet{Biazzo11a}. 
Low values of $\xi$ can lead to a large overestimate of the iron abundance. Indeed, CS~Cha is the star with the highest iron abundance, [Fe/H]=$+0.11$~dex, while the other values found by Padgett vary between $-0.26$ and $0.00$~dex. 

\subsection{Iron abundance in the Chamaeleon complex}

\begin{figure}
\begin{center}
\includegraphics[width=9.5cm]{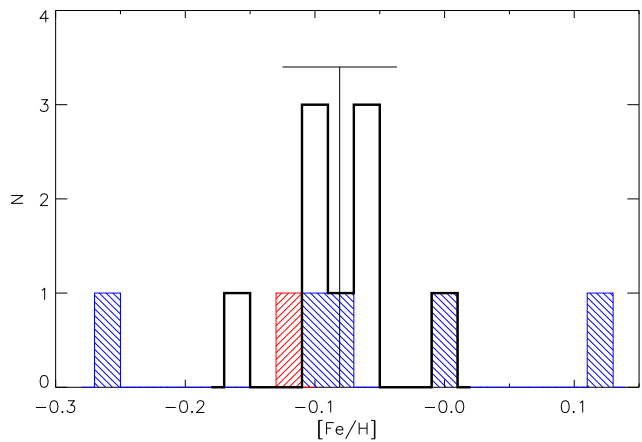}
\vspace{+.2cm}
\caption{Comparison between our [Fe/H] distribution (black histogram) with previous estimates by 
\cite{Padgett96} (blue histogram) and the single value by \cite{Biazzo12a} for Cha~II
(in red). The mean [Fe/H] for the Cha~I members and its standard deviation derived in this paper are also indicated 
by the solid line.}
 \label{fig:cha_feh}
\end{center}
\end{figure}

We now discuss the overall metallicity of the Chamaeleon complex. In Figure~\ref{fig:cha_feh} 
our [Fe/H] determination for Cha~I is compared with previous estimates 
by \cite{Padgett96}, and \cite{Biazzo12a}. Padgett analyzed five stars associated with the Cha~I dark cloud, 
while \cite{Biazzo12a} analyzed only one target in Cha~II, namely Hn~23. 
Our estimate of the average metallicity of Cha~I ([Fe/H]=$-0.08\pm0.04$) is in good agreement within the error bars with the results by \cite{Padgett96}, who derived a mean value of [Fe/H]=$-0.06\pm0.14$$\footnote{For 
consistency, we report the weighted mean of [Fe/H] as the average value of the iron 
abundance, while the error is the standard deviation around 
the mean. This determination includes the iron abundance of CS~Cha.}$.

It is also clear from Fig.~\ref{fig:cha_feh} that our [Fe/H] distribution is 
narrower than that obtained by \cite{Padgett96} 
($-0.26$ to $+0.11$~dex) and that the star-to-star variation in metallicity is smaller than the observational 
errors. Unlike Padgett's conclusion that the dispersion of [Fe/H] in Cha~I is larger 
than that of older clusters, our analysis shows that this is not the case. We believe that Padgett's results can be attributed to the uncertainties that affect the abundance analysis of young stars (e.g., low quality spectra, uncertain stellar parameters, 
high activity level, etc.) and not to a real dispersion 
in metallicity of a given SFR. Indeed, homogeneous abundance measurements have been found in other SFRs, such as the 
sub-groups of the Orion complex reported by \citet{GonzalezHernandez08}, \citet{DOrazi09b}, 
\citet{Biazzo11a,Biazzo11b} and the Taurus-Auriga association analyzed by \citet{DOrazi11}.

Furthermore, we note that the iron abundance [Fe/H]$=-0.12\pm0.14$ found in Cha~II by \cite{Biazzo12a} from the analysis of the UVES spectrum of Hn~23 strengthens our conclusion for a subsolar metallicity of the Chamaeleon complex. In order to test the consistency of the Gaia-ESO results with those of \cite{Biazzo12a} on Hn~23, we have analyzed the Gaia-ESO spectrum of the Cha~I member $\#$36, adopting the same procedure and tools (linelist, atmospheric models, etc...) used by \cite{Biazzo12a}. 
The derived atmospheric parameters, $\rm T_{\rm eff}$$=$5230$\pm$40~K, log~g$=$3.95$\pm$0.15~dex and [Fe/H]$=$$-$0.08$\pm$0.07~dex, are in excellent agreement with those produced by the Gaia-ESO Survey.

\subsection{Metallicity in nearby YOCs and SFRs}

Recently, \citet{Biazzo11a} have presented a comprehensive comparison of the
metallicity of YOCs and SFRs in the solar neighborhood (within 500~pc from the Sun). They showed that YOCs have an iron content similar to the solar value, while SFRs appear slightly more metal-poor than the Sun. In particular, and most interestingly,
no metal-rich SFRs seem to exist within this volume. However, these conclusions are based both on small number statistics (typically, 1-5 stars per region), 
and on [Fe/H] values determined from different observations and methods of analysis. Therefore, new homogeneous studies are needed to enable a more rigorous view on the metal content in nearby YOCs and SFRs to be developed. The Gaia-ESO Survey will contribute significantly to this aspect.

Our metallicity determination of Cha~I, the first SFR observed by the Gaia-ESO Survey, is in line with other metal-poor SFRs analyzed by \citet{Biazzo11a}. Similarly, Gamma Velorum, the first YOC of the Gaia-ESO Survey, with an iron content of $<$[Fe/H]$>$$=$$-$0.057$\pm$0.018~dex \citep{Spina14} is consistent with Cha~I within the errors. We have also shown that in both cases there is no dependence of [Fe/H] on effective temperature. Hence, the two regions share the same metallicity and, most importantly, the determination is based on the same methods. Although definitive conclusions will be drawn once the Gaia-ESO consortium will produce the analysis of additional regions, the initial evidence suggests that: {\it i)} there is no systematic offset between the metallicity
of YOCs and SFRs due to the analysis; {\it ii)} young clusters
can also be more metal poor than the Sun, implying that their subsolar abundance
is possibly related to their origin.

\begin{figure}
\begin{center}
\includegraphics[width=9.5cm]{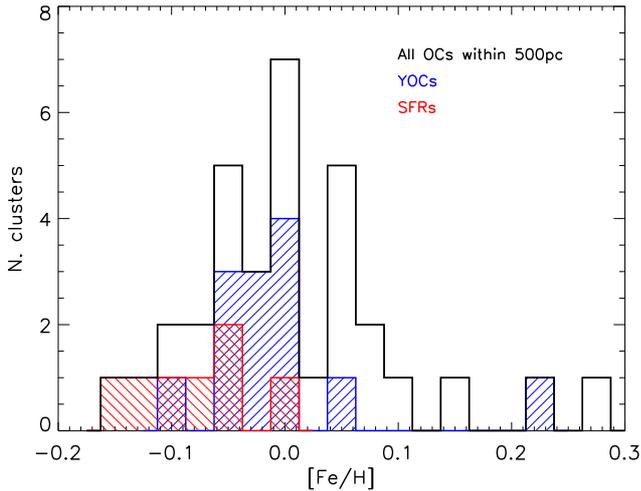}
\vspace{+.2cm}
\caption{[Fe/H] distribution of the open clusters and SFRs in the solar neighborhood within a distance of 500~pc. Adopted ages, distances and metallicity values are listed in Table~\ref{clusters_metallicity}. The red and blue colors denote the SFRs and YOCs subsamples, respectively.
}
 \label{cluster_iron_distr}
\end{center}
\end{figure}

In order to further investigate this aspect, in Fig.~\ref{cluster_iron_distr} we display the metallicity distribution of all 
the clusters in the solar neighborhood with a determination of the iron abundance based on spectra characterized by a SNR greater than 20 and a resolution greater than R$\sim$7500. These determinations are listed in Table~\ref{clusters_metallicity}.
The clusters cover a range in [Fe/H] from $-$0.20 to $+$0.27~dex, but
the youngest associations ($\lesssim$100~Myr) are generally restricted to the low metallicity values. 
Because of their young ages, these regions have not  had time to migrate through and disperse in the Galactic disk.
Thus, their metal content is representative of the present chemical pattern 
of the interstellar medium in the solar neighborhood. 
Since the chemical content and metallicity
provide a powerful tool for tagging groups of stars or associations to a common formation site \citep{Freeman02,Tabernero12,Mitschang13,Magrini14}, the result of Fig.~\ref{cluster_iron_distr} suggests these young associations may share the same origin.

In this context, it is interesting to consider the Gould Belt (GB), a structure clearly visible in the sky as a large ring of mainly O- and B- type 
stars (for a detailed discussion see \citealt{Poppel97}). 
The ring has a diameter of $\sim$1000~pc and is tilted 
by $\sim$2$0^{\circ}$ with respect to the Galactic plane. The GB is a relatively recent structure that formed between 20 and 90~Myr ago \citep{Torra00}.
 Currently, the Sun is located within the ring at $\sim$100~pc from its center. 
 The origin of the GB is still uncertain. Some studies suggest that the structure formed from the strong stellar wind originating in the central Cas-Tau OB association (e.g., \citealt{Blaauw91,Poppel97}, and references therein). 
 Other authors proposed that the GB formed from the collision of high-velocity clouds with the interstellar medium of the Galactic disk (e.g., \citealt{Comeron94}). 
 However, a combined scenario is also conceivable: the stellar feedback of massive OB stars and supernovae compressed the medium in the Galactic disk generating an expanding gaseous ring and simultaneously blowing out clumps of gas that subsequently fell back into the mid-plane of the galactic disk  \citep{Bally08}. 

\begin{figure}
\begin{center}
\includegraphics[width=9.5cm]{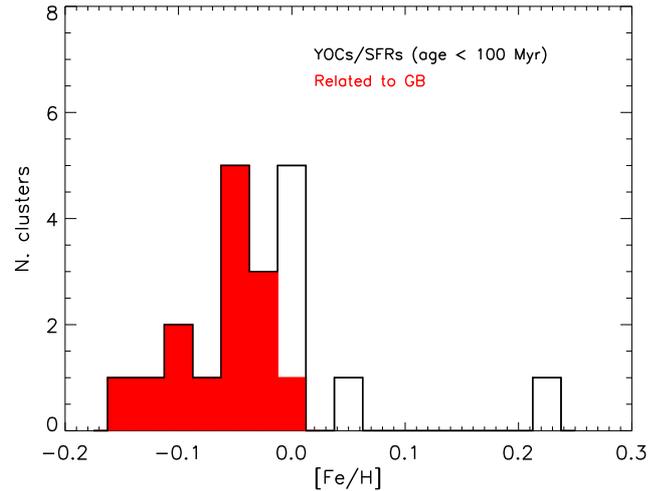}
\vspace{+.2cm}
\caption{Iron abundance distribution of the YOCs/SFRs in the solar neighborhood. The colored part of the histogram indicates the objects associated with the GB.}
 \label{yoc_sfr}
\end{center}
\end{figure}

Remarkably, the GB contains most of the SFRs and YOCs in the solar neighborhood: in the last column of Table~\ref{clusters_metallicity} 
we report the information on whether or not the cluster is associated with the GB
according to the studies by \citet{Poppel97,deZeeuw99,Elias09}. For the 
latter study, we consider as associated with the GB those clusters with a probability greater than 90~\%.
In Fig.~\ref{yoc_sfr} we plot the metallicity distribution of the clusters with an age $\leq$100~Myr, separately for clusters
associated and not associated with the GB. 
The figure and the table clearly show that a large fraction of
the nearby clusters are indeed associated with the GB and that most of them
have subsolar metallicity. Conversely, the most metal-rich clusters and SFRs
in the sample are not associated with the GB. We performed a two sample test
using ASURV survival analysis package \citep{Lavalley92} and found that the probability
that clusters associated and not associated with the GB
are drawn from the same parent population is below 0.3$\%$, or, conversely, these two distributions of metallicity are different at the 3$\sigma$ level.
These facts lead us to suggest that the SFRs and YOCs associated with the GB show a metallicity 
distinctively lower than that of the Sun and that this could offer a reasonable 
explanation for the metal-poor nature found for most of the youngest stars 
in the solar neighborhood.
 
We caution that this conclusion is tentative due to the small sample analyzed so far. For example, one might argue that metal-rich clusters
related to the GB do exist, but have not yet been observed due to the
still low number statistics. Also, we have already pointed out that most of the metallicity
determinations are heterogeneous and, in some cases, affected by large errors. 
Furthermore, uncertainties in distances and proper motions prevent a conclusive assessment of the association of the various SFRs and YOCs with the GB. Hence, additional homogeneous
and accurate information on the chemistry, the kinematics and the dynamics 
is needed to further investigate the issue.
The Gaia-ESO Survey will soon provide critical data on several other young clusters. Similarly, the Gaia mission will trace the GB structure with unprecedented accuracy, removing the uncertainty on membership and dynamical history.
If these critical measurements confirm the metal-poor nature of the clusters/SFRs associated with the GB, it will certainly
give important hints on the processes that generated the GB itself.

\begin{table*}
\tiny
\begin{center}
\begin{threeparttable}
\caption{\label{clusters_metallicity} Metallicity of open clusters in the solar Neighborhood ($\lesssim$500~pc)}
\begin{tabular}{cccccccc} 
\hline\hline 
Name & Age & Dist. & Ref. & [Fe/H] & $\#$ stars & Ref. & Gould Belt \\
 & (Myr) & (pc) & & (dex) &  & & association \\
\hline\hline 
\multicolumn{8}{c}{Star-Forming Regions} \\ \hline
ONC & 2 & 400 & 31 & -0.11$\pm$0.08 & 11 & \citet{Biazzo11a} & Y \\
Corona Australis & 3 & 138 & 46 & -0.06$\pm$0.05 & 3 & \citet{Santos08} & Y \\
Lupus & 3 & 155 & 3, 32 & -0.05$\pm$0.01 & 5 & \citet{Santos08} & Y \\
Rho Ophiuchi & 3 & 120 & 26, 33 & -0.14$\pm$0.02 & 2 & \citet{Randich14} & Y \\
Taurus & 1 & 140 & 12, 10 & -0.01$\pm$0.05 & 6 & \citet{DOrazi11} & N \\
Cha I & 2 & 160 & 30, 8 & -0.08$\pm$0.04 & 9 & this work & Y \\
Cha II & 4 & 178 & 36, 8 & -0.12 & 1 & \citet{Biazzo12a} & Y \\  \hline  \hline
\multicolumn{8}{c}{Young Open Clusters} \\ \hline
Orion OB1b & 5 & 400 & 28 & -0.05$\pm$0.05 & 5 & \citet{Biazzo11a} & Y \\
25 Ori & 10 & 330 & 24 & -0.05$\pm$0.05 & 5 & \citet{Biazzo11b} & Y \\
Sigma Ori & 4 & 360 & 23, 2 & -0.02$\pm$0.09 & 9 & \citet{GonzalezHernandez08} & Y \\
Lambda Ori & 10 & 400 & 17 & 0.01$\pm$0.01 & 5 & \citet{Biazzo11b} & Y \\
Gamma Velorum & 10 & 350 & 37 & -0.057$\pm$0.018 & 7 & \citet{Spina14} & Y \\
IC 2602 & 30 & 145 & 15, 7 & 0.00$\pm$0.01 & 8 & \citet{DOrazi09a} & N \\
IC 2391 & 55 & 149 & 15, 11, 21 & -0.01$\pm$0.02 & 7 & \citet{DOrazi09a} & N \\
IC 4665 & 25 & 385 & 34, 16 & -0.03$\pm$0.04 & 18 & \citet{Shen05} & Y \\
NGC 2451A & 57 & 188 & 25, 14 & -0.01$\pm$0.08 & 6 & \citet{Hunsch04} & N \\
Melotte 20 & 60 & 172 & 46, 40 & 0.23$\pm$0.08 & 2 & 	\citet{Gonzalez96} & N \\
Blanco 1 & 90 & 207 & 5, 40 & 0.04$\pm$0.02 & 8 & \citet{Ford05} & N \\
Upper Scorpius & 10 & 140 & 45, 35 & -0.09$\pm$0.10 & 6 & \citet{Randich14} & Y \\ 
Upper Centaurus Lupus & 16 & 140 & 1, 13 & -0.02$\pm$0.05 & 2 & \citet{Randich14} & Y \\  \hline  \hline
\multicolumn{8}{c}{Older Open Clusters} \\ \hline
Hyades & 625 & 46 & 6 & 0.11$\pm$0.01 & 3 & \citet{Carrera11} & N \\
IC 4756 & 790 & 430 & 22,41 & 0.02$\pm$0.03 & 6 & \citet{Santos09} & N \\
M34 & 200-250 & 475 & 4 & 0.07$\pm$0.04 & 9 & \citet{Schuler03} & N \\
Melotte 111 & 450 & 86 & 27, 15 & 0.06$\pm$0.10 & 22 & \citet{Gebran08} & N \\
NGC 752 & 1590 & 400 & 42 & 0.01$\pm$0.04 & 18 & \citet{Sestito04} & N \\
NGC 1901 & 400 & 400 & 29 & -0.08 & 1 & \citet{Carraro07} & N \\
NGC 2516 & 158 & 360 & 20 & 0.01$\pm$0.07 & 2 & \citet{Terndrup02} & N \\
NGC 3532 & 320 & 492 & 43 & 0.04$\pm$0.05 & 6 & \citet{Smiljanic09} & N \\
NGC 6281 & 316 & 512 & 38 & 0.05$\pm$0.06 & 2 & \citet{Smiljanic09} & N \\
NGC 6475 & 200 & 300 & 39 & 0.14$\pm$0.06 & 13 & 	\citet{Sestito03} & N \\
NGC 6633 & 600 & 376 & 18, 19 & 0.06$\pm$0.01 & 3 & \citet{Santos09} & N \\
Pleiades & 120 & 120 & 9, 40 & 0.07$\pm$0.05 & 20 & \citet{Soderblom09} & N \\
Praesepe & 794 & 182 & 25, 40 & 0.27$\pm$0.04 & 7 & \citet{Pace09} & N \\ 
\hline \hline
\end{tabular}
\begin{tablenotes}
      \tiny
      \item 1: \citet{deGeus89}; 2: \citet{Brown94}; 3: \citet{Hughes94}; 4: \citet{Jones96a}; 5: \citet{Panagi97}; 6: \citet{Perryman97}; 7: \citet{Stauffer97}; 8: \citet{Whittet97}; 9: \citet{Stauffer98}; 10: \citet{Wichmann98}; 11: \citet{Barrado99}; 12: \citet{Briceno99}; 13: \citet{deZeeuw99}; 14: \citet{Robichon99}; 15: \citet{vanLeeuwen99}; 16: \citet{Hoogerwerf00}; 17: \citet{Dolan02}; 18: \citet{Dias02}; 19: \citet{Jeffries02}; 20: \citet{Sung02}; 21: \citet{Barrado04}; 22: \citet{Salaris04}; 23: \citet{Sherry04}; 24: \citet{Briceno05}; 25: \citet{Kharchenko05}; 26: \citet{Wilking05}; 27: \citet{Casewell06}; 28: \citet{Briceno07}; 29: \citet{Carraro07}; 30: \citet{Luhman07}; 31: \citet{Menten07}; 32: \citet{Comeron08}; 33: \citet{Lombardi08}; 34: \citet{Manzi08}; 35: \citet{Preibisch08}; 36: \citet{Spezzi08}; 37: \citet{Jeffries09}; 38: \citet{Smiljanic09}; 39: \citet{Villanova09}; 40: \citet{vanLeeuwen09}; 41: \citet{Pace10}; 42: \citet{Carrera11}; 43: \citet{Clem11}; 44: \citet{SiciliaAguilar11}; 45: \citet{Pecaut12}; 46: \citet{Zuckerman12}. 
    \end{tablenotes}
\end{threeparttable}
\end{center}
\end{table*}

\section{Conclusions \label{Conclusions}}
In this paper we have used the dataset provided by the Gaia-ESO Survey to confirm the membership of Cha~I of a number of the candidate members identified by L08
and to study their metallicities. We have found that Cha~I has a slightly subsolar iron abundance, $<$[Fe/H]$>$$=-$0.08$\pm$0.04~dex, derived from nine members observed with UVES and located in different parts of the complex. The small dispersion suggests that the two stellar groups of Cha~I and the sparse population have a homogeneous metal content, as expected for a T~Tauri association such as Cha~I, isolated from other SFRs, YOCs, and OB associations.\\

The other findings can be summarized as follows:\\

i) We have confirmed the membership of Cha~I of fifteen L08 stars on the basis of the surface gravity, the presence of photospheric lithium and their position in the HRD. These stars belong to the two sub-clusters Cha~I North and South and to the sparse population around the main molecular cloud. The sample of UVES targets does not contain any new member of Cha~I. \\
\\
ii) Our determination of the metallicity allows us to better constrain the
[Fe/H] distribution and the average [Fe/H] value of Cha~I. The mean value of [Fe/H]
agrees reasonably well with that obtained by \citet{Padgett96}, but our
dispersion is much smaller. Also, the metallicity of Cha~I is similar to that
of Cha~II derived by \citet{Biazzo12a}. This result indicates that the whole Chamaeleon complex is more metal-poor than the Sun.\\
\\
iii) We speculate that the metallicity of Cha~I is similar to that of other SFRs in the solar neighborhood. The metal-poor nature of these young environments could be the result of a common and widespread star formation episode that involved the Gould Belt and that gave birth to most of the SFRs and YOCs in the solar vicinity. 

Our study not only reinforces the hints that the youngest stars in the Solar neighborhood are poorer of metals than the Sun itself, but it also shows the great potential of the Gaia-ESO Survey on this type of scientific research, thanks to its homogeneous analysis and the rich statistics.

\begin{acknowledgement}
Based on data products from observations made with ESO Telescopes at the La Silla Paranal Observatory under programme ID 188.B-3002.
This work was partly supported by the European Union FP7 programme through ERC grant number 320360 and by the Leverhulme Trust through grant RPG-2012-541.
We acknowledge the support from INAF and Ministero dell' Istruzione, dell' Universit\`a' e della Ricerca (MIUR) in the form of the grant ``Premiale VLT 2012''.
The results presented here benefit from discussions held during the Gaia-ESO workshops and conferences supported by the ESF (European Science Foundation) through the GREAT Research Network Programme.
We acknowlodge the use of the ASURV astronomy survival analysis package \citep{Lavalley92} which is freely available from http://www.astro.psu.edu/statcodes/asurv . E.D.M., S.G.S. and V.Zh.A. acknowledge the support from the Funda\c{c}\~ao para a Ci\^encia e Tecnologia, FCT (Portugal) in the form of the fellowships SFRH/BPD/76606/2011, SFRH/BPD/47611/2008 and SFRH/BPD/70574/2010 from the FCT (Portugal), respectively.
We also acknowledge the support of the French Agence Nationale de la
Recherche, under contract ANR-2010-BLAN-0508-01OTP, and the ``Programme
National de Cosmologie et Galaxies'' (PNCG) of CNRS/INSU, France.
J.I.G.H. acknowledges financial support from the Spanish Ministry project MINECO AYA2011-29060, 
and also from the Spanish Ministry of Economy and Competitiveness (MINECO) under the 2011 Severo Ochoa Program MINECO SEV-2011-0187.
H.M.T. and D.M. acknowledges financial support from the Spanish Ministry of Economy and Competitiveness (MINECO) 
project AYA2011-30147-C03-02.
T.B. was funded by grant No. 621-2009-3911 from The Swedish Research Council.

\end{acknowledgement}

\bibliographystyle{aa} 
\bibliography{bibliography}

\onllongtab{

\longtab{1}{ 
\begin{threeparttable}
\begin{longtable}{cccccccc} 
\caption{\label{members}The target stars}\\ 
\hline\hline 
ID & CNAME & RA & DEC & $J_{\rm 2MASS}$ & SNR & Membership & Tight \\
 & & (J2000) & (J2000) & (mag) & & (L08) &  system* \\
\hline 
\endfirsthead 
\caption{continued.}\\ 
\hline\hline 
ID & CNAME & RA & DEC & $J_{\rm 2MASS}$ & SNR & Membership & Tight \\
 & & (J2000) & (J2000) & (mag) & & (L08) &  system* \\
\hline 
\endhead 
\hline 
\endfoot 
1 & 10554858-7651504 & 10 55 48.58 & -76 51 50.4 & 9.58 & 103 & --- & --- \\
2 & 10555973-7724399 & 10 55 59.73 & -77 24 39.9 & 10.78 & 6 & Y & N \\
3 & 10564115-7744292 & 10 56 41.15 & -77 44 29.2 & 7.72 & 74 & --- & --- \\
4 & 10574797-7617429 & 10 57 47.97 & -76 17 42.9 & 7.84 & 101 & --- & --- \\
5 & 10585418-7743115 & 10 58 54.18 & -77 43 11.5 & 10.79 & 48 & --- & ---\\
6 & 10590108-7722407 & 10 59 01.08 & -77 22 40.7 & 10.14 & 26 & Y & N \\
7 & 10593816-7822421 & 10 59 38.16 & -78 22 42.1 & 7.74 & 73 & --- & --- \\
8 & 11010007-7738516 & 11 01 00.07 & -77 38 51.6 & 9.49 & 45 & --- & --- \\
9 & 11012887-7539520 & 11 01 28.87 & -75 39 52.0 & 7.02 & 274 & --- & --- \\
10 & 11020524-7525093 & 11 02 05.24 & -75 25 09.3 & 7.27 & 99 & --- & --- \\
11 & 11022491-7733357 & 11 02 24.91 & -77 33 35.7 & 9.10 & 77 & Y & N \\
12 & 11033599-7628242 & 11 03 35.99 & -76 28 24.2 & 8.71 & 62 & --- & --- \\
13 & 11034945-7700101 & 11 03 49.45 & -77 00 10.1 & 8.58 & 22 & --- & --- \\
14 & 11044460-7706240 & 11 04 44.60 & -77 06 24.0 & 9.60 & 34 & --- & --- \\
15 & 11045100-7625240 & 11 04 51.00 & -76 25 24.0 & 10.54 & 27 & Y & N \\
16 & 11053303-7700120 & 11 05 33.03 & -77 00 12.0 & 10.73 & 12 & --- & --- \\
17 & 11055780-7607489 & 11 05 57.80 & -76 07 48.9 & 7.15 & 296 & Y & N \\
18 & 11060511-7511454 & 11 06 05.11 & -75 11 45.4 & 7.72 & 129 & --- & --- \\
19 & 11064510-7727023 & 11 06 45.10 & -77 27 02.3 & 10.18 & 9 & Y & N \\
20 & 11065856-7713326 & 11 06 58.56 & -77 13 32.6 & 7.70 & 47 & --- & --- \\
21 & 11075588-7727257 & 11 07 55.88 & -77 27 25.7 & 9.22 & 38 & Y & Y \\
22 & 11080148-7742288 & 11 08 01.48 & -77 42 28.8 & 8.70 & 80 & Y & Y \\
23 & 11080412-7513273 & 11 08 04.12 & -75 13 27.3 & 8.92 & 65 & --- & --- \\
24 & 11082577-7648315 & 11 08 25.77 & -76 48 31.5 & 7.88 & 86 & --- & --- \\
25 & 11084041-7756310 & 11 08 40.41 & -77 56 31.0 & 9.90 & 69 & --- & --- \\
26 & 11085231-7743329 & 11 08 52.31 & -77 43 32.9 & 12.85 & $<$5 & --- & --- \\
27 & 11085326-7519374 & 11 08 53.26 & -75 19 37.4 & 9.74 & 42 & Y & --- \\
28 & 11091172-7729124 & 11 09 11.72 & -77 29 12.4 & 9.93 & 39 & Y & Y \\
29 & 11091769-7627578 & 11 09 17.69 & -76 27 57.8 & 10.00 & 58 & Y & Y \\
30 & 11092378-7623207 & 11 09 23.78 & -76 23 20.7 & 10.44 & 93 & Y & N \\
31 & 11095119-7658568 & 11 09 51.19 & -76 58 56.8 & 9.52 & 47 & --- & --- \\
32 & 11100704-7629377 & 11 10 07.04 & -76 29 37.7 & 9.91 & 59 & Y & Y \\
33 & 11111333-7731178 & 11 11 13.33 & -77 31 17.8 & 8.02 & 42 & --- & --- \\
34 & 11112801-7749213 & 11 11 28.01 & -77 49 21.3 & 8.77 & 56 & --- & --- \\
35 & 11114632-7620092 & 11 11 46.32 & -76 20 09.2 & 9.11 & 147 & Y & N \\
36 & 11124268-7722230 & 11 12 42.68 & -77 22 23.0 & 8.65 & 169 & Y & Y \\
37 & 11124299-7637049 & 11 12 42.99 & -76 37 04.9 & 10.06 & 92 & Y & N \\
38 & 11135757-7818460 & 11 13 57.57 & -78 18 46.0 & 14.93 & $<$5 & --- & --- \\
39 & 11140585-7729058 & 11 14 05.85 & -77 29 05.8 & 8.31 & 45 & --- & --- \\
40 & 11140941-7714492 & 11 14 09.41 & -77 14 49.2 & 7.37 & 121 & --- & --- \\
41 & 11141568-7738326 & 11 14 15.68 & -77 38 32.6 & 10.72 & 58 & --- & --- \\
42 & 11142964-7707063 & 11 14 29.64 & -77 07 06.3 & 9.69 & 64 & --- & --- \\
43 & 11143515-7539288 & 11 14 35.15 & -75 39 28.8 & 9.29 & 84 & --- & --- \\
44 & 11170509-7538518 & 11 17 05.09 & -75 38 51.8 & 7.35 & 137 & --- & --- \\
45 & 11182024-7621576 & 11 18 20.24 & -76 21 57.6 & 9.79 & 40 & Y & Y \\
46 & 11213017-7616098 & 11 21 30.17 & -76 16 09.8 & 9.59 & 65 & --- & --- \\
47 & 11252677-7553273 & 11 25 26.77 & -75 53 27.3 & 8.50 & 127 & --- & --- \\
48 & 11291261-7546263 & 11 29 12.61 & -75 46 26.3 & 9.82 & 38 & Y & --- \\ \hline \hline
\end{longtable} 
\begin{tablenotes}
      \tiny
      \item * We consider ``tight systems'' only the multiple systems with a separation $\leq$2$\arcsec$ between the components, assuming the determinations from \citet{Lafreniere08}.
    \end{tablenotes}
\end{threeparttable}
}

\longtab{2}{ 
\tiny
\begin{threeparttable}
\begin{longtable}{ccccccccccccc} 
\caption{\label{67parameters} Stellar parameters of the 48 UVES stars.}\\ 
\hline\hline 
ID & RV & v~sin{\it i} & $\rm T_{\rm eff}$ & log g & [Fe/H] & EW(Li) & $L_{bol}$ & bin. & log g & Li & HRD & Final \\
 & (km/s) & (km/s) & (K) & (dex) & (dex) & m$\AA$ & ($L_{\sun}$) & & mem. & mem. & mem. & mem. \\
\hline 
\endfirsthead 
\caption{continued.}\\ 
\hline\hline 
ID & RV & v~sin{\it i} & $\rm T_{\rm eff}$ & log g & [Fe/H] & EW(Li) & $L_{bol}$ & bin. & log g & Li & HRD & Final \\
 & (km/s) & (km/s) & (K) & (dex) & (dex) & m$\AA$ & ($L_{\sun}$) & & mem. & mem. & mem. & mem. \\
\hline 
\endhead 
\hline 
\endfoot 
1 & 44.0$\pm$0.6 & ... & ... & ... & ... & $<$15 & ... & N & --- & --- & --- & --- \\
2 & ... & ... & 3640$\pm$300 & 4.40$\pm$0.20 & -0.10$\pm$0.20 & 406$\pm$51 & 0.38$_{-0.09}^{+0.15}$ & N & Y & Y & Y & Y \\
3 & -1.9$\pm$0.6 & 0.8$\pm$0.9 & 4384$\pm$51 & 2.02$\pm$0.18 & -0.33$\pm$0.08 & 104$\pm$1 & ... & N & N & --- & --- & N \\
4 & 4.8$\pm$0.6 & 2.1$\pm$1.4 & 5110$\pm$49 & 3.08$\pm$0.16 & 0.19$\pm$0.12 & 15$\pm$10 & ... & N & N & --- & --- & N \\
5 & 10.4$\pm$0.6 & 0.5$\pm$0.5 & 4701$\pm$60 & 2.99$\pm$0.17 & 0.13$\pm$0.10 & $<$15 & ... & N & N & --- & --- & N \\
6 & 15.3$\pm$0.6 & 8.8$\pm$0.8 & 4135$\pm$125 & 4.63$\pm$0.13 & -0.09$\pm$0.13 & 569$\pm$18 & 0.79$_{-0.19}^{+0.26}$ & N & Y & Y & Y & Y \\
7 & 10.0$\pm$0.6 & 0.5$\pm$0.5 & 4464$\pm$55 & 2.25$\pm$0.20 & -0.09$\pm$0.12 & $<$10 & ... & N & N & --- & --- & N \\
8 & 108.1$\pm$0.2 & 2.1$\pm$1.9 & 3958$\pm$63 & 1.85$\pm$0.49 & -0.31$\pm$0.15 & 317$\pm$5 & ... & N & N & --- & --- & N \\
9 & ... & ... & ...  & ... & ... & $<$2 & ... & N & --- & --- & --- & --- \\
10 & -28.1$\pm$0.6 & 1.2$\pm$1.5 & 4253$\pm$67 & 1.77$\pm$0.28 & -0.23$\pm$0.08 & $<$5 & ... & N & N & --- & --- & N \\
11 & 14.2$\pm$0.6 & 14.1$\pm$1.2 & 4656$\pm$193 & 4.28$\pm$0.52 & -0.16$\pm$0.11 & 571$\pm$5 & 1.74$_{-0.53}^{+0.71}$ & N & Y & Y & Y & Y \\
12 & 11.9$\pm$0.6 & 1.0$\pm$1.2 & 4242$\pm$80 & 1.85$\pm$0.34 & -0.22$\pm$0.06 & 35$\pm$2 & ... & N & N & --- & --- & N \\
13 & -41.5$\pm$0.6 & 0.5$\pm$0.5 & 3865$\pm$72 & 1.54$\pm$0.12 & 0.03$\pm$0.11 & $<$30 & ... & N & N & --- & --- & N \\
14 & 28.9$\pm$0.6 & 0.6$\pm$0.6 & 4411$\pm$60 & 1.99$\pm$0.18 & -0.30$\pm$0.07 & $<$10 & ... & N & N & --- & --- & N \\
15 & 13.5$\pm$0.6 & 13.8$\pm$1.2 & 4571$\pm$333 & 4.44$\pm$0.16 & 0.00$\pm$0.16 & 575$\pm$6 & 0.50$_{-0.19}^{+0.29}$ & N & Y & Y & Y & Y \\
16 & 8.9$\pm$0.6 & 2.2$\pm$1.7 & 4631$\pm$417 & 4.39$\pm$0.35 & -0.12$\pm$0.12 & 30$\pm$10 & ... & N & Y & N & --- & N \\
17 & ... & ... & ... & ... & ... & $<$5 & ... & N & --- & --- & --- & --- \\
18 & 77.1$\pm$0.6 & 3.2$\pm$1.6 & 5003$\pm$39 & 2.69$\pm$0.16 & -0.33$\pm$0.17 & $<$5 & ... & N & N & --- & --- & N \\
19 & 15.4$\pm$0.6 & 19.3$\pm$2.5 & 4343$\pm$147 & 4.56$\pm$0.15 & -0.07$\pm$0.13 & 797$\pm$32 & 1.10$_{-0.30}^{+0.40}$ & N & Y & Y & Y & Y \\
20 & -8.6$\pm$0.2 & 0.8$\pm$1.0 & 3777$\pm$85* & 1.44$\pm$0.20* & 0.02$\pm$0.11* & 212$\pm$45 & ... & N & N & --- & --- & N \\
21 & 17.5$\pm$0.6 & 9.1$\pm$1.4 & 4651$\pm$174 & 4.10$\pm$0.57 & -0.25$\pm$0.23 & 499$\pm$3 & 5.39$_{-1.59}^{+2.08}$ & N & Y & Y & Y & Y \\
22 & 4.5$\pm$0.6 & 23.3$\pm$4.6 & 4327$\pm$131 & 4.44$\pm$0.12 & -0.10$\pm$0.14 & 555$\pm$13 & ... & Y & --- & --- & --- & --- \\
23 & -15.6$\pm$0.6 & 0.5$\pm$0.6 & 4436$\pm$91 & 2.15$\pm$0.18 & -0.11$\pm$0.10 & 36$\pm$9 & ... & N & N & --- & --- & N \\
24 & 15.9$\pm$0.6 & 0.5$\pm$0.5 & 4485$\pm$53 & 2.22$\pm$0.16 & -0.01$\pm$0.16 & $<$5 & ... & N & N & --- & --- & N \\
25 & -6.5$\pm$0.6 & 29.9$\pm$6.2 & 6378$\pm$183 & 3.94$\pm$0.17 & -0.01$\pm$0.12 & 97$\pm$15 & ... & N & Y & HCM & N & N \\
26 & ... & ... & ... & ...  & ... & $<$5 & ... & N & --- & --- & --- & --- \\
27 & 50.3$\pm$0.6 & ... & ... & ... & ... & 428$\pm$8 & ... & Y & --- & --- & --- & --- \\
28 & 14.8$\pm$0.6 & 5.0$\pm$1.6 & 4175$\pm$381 & 4.38$\pm$0.60 & -0.45$\pm$0.44 & 654$\pm$15 & 0.74$_{-0.25}^{+0.49}$ & N & Y & Y & Y & Y \\
29 & 15.1$\pm$0.6 & 14.8$\pm$1.2 & 4524$\pm$103 & 4.21$\pm$0.58 & -0.14$\pm$0.12 & 567$\pm$10 & 1.41$_{-0.35}^{+0.44}$ & N & Y & Y & Y & Y \\
30 & ... & ... & 3990$\pm$123 & 4.64$\pm$0.13 & -0.09$\pm$0.13 & 547$\pm$15 & 1.21$_{-0.28}^{+0.34}$ & N & Y & Y & Y & Y \\
31 & 35.1$\pm$0.6 & 0.5$\pm$0.5 & 5755$\pm$76 & 4.28$\pm$0.19 & 0.49$\pm$0.15 & $<$10 & ... & N & Y & N & --- & N \\
32 & 14.0$\pm$0.6 & 5.1$\pm$1.2 & 4697$\pm$120 & 4.37$\pm$0.43 & -0.32$\pm$0.27 & 563$\pm$6 & 1.75$_{-0.46}^{+0.57}$ & N & Y & Y & Y & Y \\
33 & 21.6$\pm$0.6 & 0.5$\pm$0.5 & 4794$\pm$43 & 3.15$\pm$0.14 & 0.21$\pm$0.16 & $<$10 & ... & N & N & --- & --- & N \\
34 & -28.2$\pm$0.6 & 0.7$\pm$0.8 & 3928$\pm$118 & 1.43$\pm$0.20 & -0.11$\pm$0.12 & 20$\pm$10 & ... & N & N & --- & --- & N \\
35 & 16.2$\pm$0.6 & 24.2$\pm$1.4 & 4617$\pm$207 & 4.50$\pm$0.17 & -0.06$\pm$0.14 & 537$\pm$15 & 2.54$_{-0.80}^{+1.08}$ & N & Y & Y & Y & Y \\
36 & 14.2$\pm$0.6 & 8.6$\pm$0.8 & 5239$\pm$70 & 4.23$\pm$0.34 & -0.08$\pm$0.08 & 369$\pm$4 & 3.87$_{-0.86}^{+1.04}$ & N & Y & Y & Y & Y \\
37 & 14.3$\pm$0.6 & 12.2$\pm$0.9 & 4706$\pm$220 & 4.27$\pm$0.42 & -0.06$\pm$0.11 & 461$\pm$5 & 0.77$_{-0.25}^{+0.33}$ & N & Y & Y & Y & Y \\
38 & ... & ... & ... & ... & ... & $<$5 & ... & N & --- & --- & --- & --- \\
39 & -3.7$\pm$0.6 & 0.7$\pm$0.8 & 3972$\pm$103 & 1.55$\pm$0.24 & 0.00$\pm$0.10 & $<$28 & ... & N & N & --- & --- & N \\
40 & -24.5$\pm$0.6 & 1.5$\pm$1.7 & 4105$\pm$65 & 1.59$\pm$0.21 & -0.29$\pm$0.07 & $<$10 & ... & N & N & --- & --- & N \\
41 & 14.8$\pm$0.6 & 99.4$\pm$10.5 & 7075$\pm$102 & 4.16$\pm$0.14 & -0.09$\pm$0.13 & $<$50 & ... & N & Y & HCM & N & N \\
42 & -17.0$\pm$0.6 & 1.9$\pm$1.6 & 5968$\pm$59 & 4.32$\pm$0.11 & 0.03$\pm$0.04 & 23$\pm$3 & ... & N & Y & N & --- & N \\
43 & -9.1$\pm$0.6 & 0.8$\pm$0.9 & 4469$\pm$43 & 2.05$\pm$0.17 & -0.26$\pm$0.08 & $<$10 & ... & N & N & --- & --- & N \\
44 & 7.0$\pm$0.6 & 0.5$\pm$0.5 & 4882$\pm$56 & 2.80$\pm$0.15 & 0.08$\pm$0.14 & 15$\pm$5 & ... & N & N & --- & --- & N \\
45 & 13.8$\pm$0.6 & 8.3$\pm$1.2 & 4465$\pm$209 & 4.25$\pm$0.58 & -0.19$\pm$0.21 & 529$\pm$7 & 0.84$_{-0.26}^{+0.37}$ & N & Y & Y & Y & Y \\
46 & -30.4$\pm$0.6 & 1.0$\pm$1.1 & 4917$\pm$50 & 2.86$\pm$0.12 & 0.05$\pm$0.08 & $<$10 & ... & N & N & --- & --- & N \\
47 & -5.8$\pm$0.6 & 19.0$\pm$2.1 & 6380$\pm$207 & 4.04$\pm$0.22 & -0.01$\pm$0.15 & $<$10 & ... & N & Y & N & --- & N \\
48 & 15.2$\pm$0.6 & 20.8$\pm$1.2 & 4818$\pm$96 & 4.50$\pm$0.15 & -0.06$\pm$0.14 & 463$\pm$7 & 1.14$_{-0.28}^{+0.33}$ & N & Y & Y & Y & Y \\ \hline
\end{longtable} 
\begin{tablenotes}
      \tiny
      \item * For object $\#$20 the stellar parameters do not come from GESviDR1Final catalog, but they are only computed by one node.
    \end{tablenotes}
\end{threeparttable}
}
}

\end{document}